\documentclass[11pt]{article}

\usepackage[dvips]{graphicx}
\usepackage{epsf}  
\usepackage{epsfig}  

\input pubboard/babarsym
\RequirePackage{xspace}

\newcommand{\pvec}{{\bf p}}
\newcommand{\bfm}{\bf\boldmath}

\newcommand{\calB}{\ensuremath{{\cal B}}}

\newcommand{\prodB}{\ensuremath{\prod\calB_i}}
\newcommand{\bfemsix}{\ensuremath{\calB(10^{-6})}}
\newcommand{\timesix}{\ensuremath{\times10^{-6}}}

\newcommand{\DE}{\ensuremath{\Delta E}}

\newcommand{\xf}{\ensuremath{{\cal F}}}
\newcommand{\hel}{\ensuremath{{\cal H}}}
\newcommand{\thetaT}{\ensuremath{\theta_{\rm T}}}
\newcommand{\costhr}{\ensuremath{\cos\thetaT}}

\newcommand\etal{{\it et al.}}
\newcommand{\half}{\ensuremath{{1\over2}}}

\newcommand{\bfig}{\begin{figure}[htbpc!]}
\newcommand{\efig}{\end{figure}}
\newcommand\bef{\begin{figure}}
\newcommand\edf{\end{figure}}
\newcommand\dbline{\noalign{\vskip 0.10truecm\hrule}\noalign{\vskip 2pt}\noalign{\hrule\vskip 0.10truecm}}
\providecommand{\tbline}{\noalign{\vskip 0.05truecm\hrule\vskip0.05truecm}}

\newcommand\beq{\begin{equation}}
\newcommand\eeq{\end{equation}}
\newcommand\bear{\begin{array}}
\newcommand\enar{\end{array}}
\newcommand\beqa{\begin{eqnarray}}
\newcommand\eeqa{\end{eqnarray}}
\newcommand\ben{\begin{enumerate}}
\newcommand\een{\end{enumerate}}

\newcommand{\UfourS}{\ensuremath{\Upsilon(4S)}}

\newcommand{\omtoppp}{\ensuremath{{\omega\ra\pip\pim\piz}}}

\newcommand{\fomegaKp}{\ensuremath{\omega K^+}}
\newcommand{\omegaKp}{\ensuremath{\Bp\ra\fomegaKp}}
\newcommand{\BomegaKp}{\ensuremath{\calB(\omegaKp)}}

\newcommand{\fomegaKz}{\ensuremath{\omega K^0}}
\newcommand{\fomegaKs}{\ensuremath{\omega\KS}}
\newcommand{\omegaKz}{\ensuremath{\Bz\ra\fomegaKz}}
\newcommand{\omegaKs}{\ensuremath{\Bz\ra\fomegaKs}}
\newcommand{\BomegaKz}{\ensuremath{\calB(\omegaKz)}}
\newcommand{\romegaKz}{\ensuremath{5.9^{+1.0}_{-0.9}\pm xx}}
\newcommand{\RomegaKz}{\ensuremath{(\romegaKz)\times 10^{-6}}}
\newcommand{\SomegaKz}{\ensuremath{0.50^{+0.32}_{-0.36}\pm zz}}
\newcommand{\ComegaKz}{\ensuremath{-0.56^{+0.27}_{-0.25}\pm zz}}
\newcommand{\somegaKz}{\ensuremath{xx}}

\renewcommand{\romegaKz}{\ensuremath{5.9\pm1.0\pm 0.4}}
\renewcommand{\somegaKz}{\ensuremath{8.6}}
\renewcommand{\SomegaKz}{\ensuremath{0.50^{+0.34}_{-0.38}\pm0.02}}
\renewcommand{\ComegaKz}{\ensuremath{-0.56^{+0.29}_{-0.27}\pm0.03}}
\providecommand{\dt}{\deltat}
\providecommand{\tcp}{\mbox{$t_{CP}$}}
\newcommand{\ttag}{\ensuremath{t_{\rm tag}}}
\newcommand{\bflav}{\ensuremath{B_{\rm flav}}}
\providecommand{\sigdt}{\ensuremath{\sigma_{\deltat}}}

\newcommand{\BABARPubYear}    {05}

\newcommand{\BABARConfNumber} {01}
\newcommand{\SLACPubNumber} {11039}
\newcommand{\LANLNumber} {0503018}

\long\def\inst#1{\par\nobreak\kern 4pt\nobreak
    {\it #1}\par\vskip 10pt plus 3pt minus 3pt}

\begin{document}
{\pagestyle{empty}


\begin{flushright}
\babar-CONF-\BABARPubYear/\BABARConfNumber \\
SLAC-PUB-\SLACPubNumber \\
hep-ex/\LANLNumber \\
March 2005 \\
\end{flushright}

\par\vskip 3cm

\begin{center}
\Large \bf Measurement of Branching Fraction and {\bfm \CP }-Violating 
Asymmetry for {\bfm \omegaKs }
\end{center}
\bigskip

\begin{center}
\large The \babar\ Collaboration\\
\mbox{ }\\
\today
\end{center}
\bigskip \bigskip

\begin{center}
\large \bf Abstract
\end{center}

We present a preliminary measurement of the branching fraction and 
\CP-violating parameters $S$ and $C$ for the decay \omegaKs. The data 
sample corresponds to $232\times 10^6$ \BB\ pairs produced
from \epem\ annihilation at the \UfourS\ resonance. 
We measure $\BomegaKz=\RomegaKz$, $S=\SomegaKz$ and $C=\ComegaKz$.

\vfill
\begin{center}
Presented at the $40^{\rm th}$ Rencontres de Moriond on\\
Electroweak Interactions and Unified Theories,\\
5---12 March 2005, La Thuile, Vall\'ee d'Aoste, Italy
\end{center}

\vspace{1.0cm}
\begin{center}
{\em Stanford Linear Accelerator Center, Stanford University, 
Stanford, CA 94309} \\ \vspace{0.1cm}\hrule\vspace{0.1cm}
Work supported in part by Department of Energy contract DE-AC03-76SF00515.
\end{center}

} 
\newpage
\begin{center}
\small

The \babar\ Collaboration,
\bigskip

B.~Aubert,
R.~Barate,
D.~Boutigny,
F.~Couderc,
Y.~Karyotakis,
J.~P.~Lees,
V.~Poireau,
V.~Tisserand,
A.~Zghiche
\inst{Laboratoire de Physique des Particules, F-74941 Annecy-le-Vieux, France }
E.~Grauges
\inst{IFAE, Universitat Autonoma de Barcelona, E-08193 Bellaterra, Barcelona, Spain }
% ...

\end{center}\newpage

\section{Introduction}
\label{sec:Introduction}
Evidence for the decay \omegaKz\ was first seen by CLEO \cite{CLEO} with
a significance of 3.9 standard deviations ($\sigma$).  The decay was 
observed about a year ago with a sample of 89 million \BB\ pairs by 
\babar\ \cite{BABAR} with a significance 
of more than $7\sigma$; the branching fraction was measured to be
$\BomegaKz=(5.9^{+1.6}_{-1.3}\pm0.5)\timesix$.
More recently Belle has published evidence (significance $3.2\sigma$) 
for this decay mode with a sample of 85 million \BB\ pairs \cite{Belle},
finding a branching fraction of $(4.0^{+1.9}_{-1.6}\pm0.5)\timesix$.
Belle also has a preliminary study with a sample of 275 million \BB\ pairs 
of the time-dependence of the decay with a
sample of about 30 events \cite{BelleTD}, though no branching fraction
is reported.

The world average branching fraction, $(5.5^{+1.2}_{-1.1})\timesix$
\cite{HFAG}, is somewhat larger than older theoretical predictions
\cite{AKL, Cheng, DDO} and a more recent prediction using QCD factorization 
\cite{BN}.  A very recent paper \cite{CCS} finds an enhancement of the QCD 
factorization prediction by more than a factor of two due to
final-state interactions, in excellent agreement with the world average.
A phenomenological fit that uses SU(3) flavor symmetry and all
available measurements of pseudoscalar-vector (PV) decays (branching fraction 
and \CP asymmetry measurements for more than 30 charmless decay modes) 
finds a branching 
fraction of $(5.3^{+0.8}_{-0.4})\timesix$ for this decay \cite{Chiang}.

\begin{figure}[htbp]
\begin{center}
\includegraphics[scale=0.8]{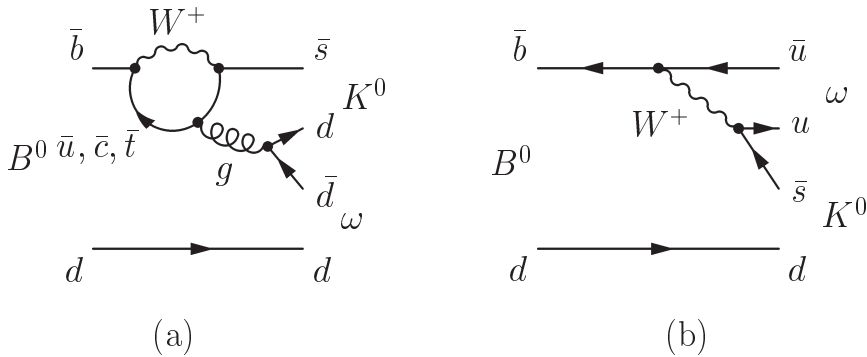}
\caption{Feynman diagrams for the decay \omegaKz:
(a) penguin diagram and (b) Cabibbo- and color-suppressed tree diagram.}
\end{center}
  \label{fig:feyn}
\end{figure}

In this paper we report improved branching fraction results for this
decay as well as a measurement of the \CP\ asymmetry parameters.  In
the Standard Model (SM), this
decay is expected to proceed primarily through a penguin (loop) diagram
as shown in Fig. \ref{fig:feyn}a, though a Cabibbo- and color-suppressed
tree diagram is also possible (Fig. \ref{fig:feyn}b).  Neglecting the
suppressed amplitude, these decay modes have the same weak phase as the 
charmonium $K^0$ decays \cite{s2b} which proceed through the
Cabibbo-Kobayashi-Maskawa (CKM) favored $b\to c\bar{c}s$ amplitude.  Thus 
the time-dependent asymmetry measurement for the decay \omegaKz\ would yield 
the same value of \stwob\ as for the charmonium $K^0$ decays \cite{lonsoni}.
Tests of this equality have been made from similar \Bz\ decays which are 
expected to be dominated by penguin amplitudes such as those to the
charmless final states $\phi\Kz$, $\etapr\Kz$, $K^+K^-\Kz$, $\piz\Kz$ and 
$f_0(980)\Kz$ \cite{HFAGUT}.

Additional higher-order amplitudes and non-SM amplitudes carrying different 
weak phases would lead to
differences between the measurements of the time-dependent \CP\ violating 
parameter in these rare decay modes and in the charmonium \Kz\ decays.  
The recent calculation involving final-state interactions that was
mentioned above \cite{CCS} predicts an increase in \stwob\ of about
0.10 for \omegaKz\ due to the color-suppressed amplitudes, though this 
increase is nullified when final-state interactions are included.

\section{The \babar\ Detector and Dataset}
\label{sec:babar}

The results presented here are based on data collected with the \babar\
detector~\cite{BABARNIM} at the PEP-II asymmetric \epem\
collider located at the Stanford Linear Accelerator Center.  We use a 
data sample with an integrated luminosity of 211~fb$^{-1}$ recorded
at the $\Upsilon (4S)$ resonance (center-of-mass energy $\sqrt{s}=10.58\
\gev$).  This corresponds to 232 million \BB\ pairs.
The asymmetric beam configuration in the laboratory frame
provides a boost of $\beta\gamma = 0.56$ to the $\Upsilon(4S)$.

Charged particles from the \epem\ interactions are detected and their
momenta measured by a combination of a vertex tracker (SVT) consisting
of five layers of double-sided silicon microstrip detectors and a
40-layer central drift chamber, both operating in the 1.5-T magnetic
field of a superconducting solenoid. We identify photons and electrons 
using a CsI(Tl) electromagnetic calorimeter (EMC).
Further charged particle identification (PID) is provided by the average energy
loss ($dE/dx$) in the tracking devices and by an internally reflecting
ring imaging Cherenkov detector (DIRC) covering the central region.
The flux return of the solenoid is composed of multiple layers of iron
and resistive plate chambers for the identification of muons and long-lived
neutral hadrons.

\section{Time-dependent Analysis}
\label{sec:dt}
From a \BB\ pair we reconstruct a \Bz\ or \Bzb\ decaying into the \CP\ 
eigenstate \omegaKs\ ($B_{CP}$).  We also reconstruct the vertex of the other 
$B$ meson ($B_{\rm tag}$) and identify its flavor.  The time difference 
$\dt\equiv\tcp -\ttag$, where $\tcp$ and $\ttag$ are the proper decay times 
of the signal and tagged $B$ mesons, respectively, is obtained from the measured 
distance between the $B_{CP}$ and $B_{\rm tag}$ decay vertices and from the 
boost ($\beta \gamma =0.56$) of the $\Upsilon(4S)$ system. 
The distribution of \dt\ without detector resolution effects is:
\begin{equation}
F( {\dt}) = \frac{e^{-\left|\dt\right|/\tau}}{4\tau} \left\{1 \mp {\Delta w} 
\pm (1 -2w) \left[ S\sin(\deltamd\dt) - C\cos(\deltamd\dt)\right]\right\},
\label{fplusminus}
\end{equation}
where the upper (lower) sign denotes a decay accompanied by a \Bz\ (\Bzb) tag, 
$\tau$ is the \Bz\ lifetime \cite{PDG2004}, \deltamd\ is the mixing frequency,
and the mistag parameters $w$ and $\Delta w$ are respectively the average and 
difference of the probabilities that a true \Bz\ (\Bzb ) meson is tagged 
as \Bzb\ (\Bz ). The tagging algorithm \cite{s2b} has seven mutually exclusive 
tagging categories of differing purities (including one for untagged events 
that we retain for the branching fraction determination).  Separate neural 
networks are trained to identify primary leptons, kaons, soft pions from 
$D^*$ decays, and high-momentum charged particles from \B\ decays.  Each 
event is assigned to one of these categories based on the estimated mistag 
probability and on the source of tagging information.  The measured analyzing 
power, equal to the reconstruction efficiency times $(1-2w)^2$ summed over all
categories, is 
$(30.5\pm 0.6)\%$; this is determined from a large sample of $B$-decays to fully 
reconstructed flavor eigenstates (\bflav).  The parameter $C$ measures direct 
\CP\ violation.  If $C=0$, then $S=\stwob$ aside from the 
corrections discussed in the Introduction.

\section{Event Selection and Analysis Method}
To establish the event selection criteria, we use Monte Carlo (MC) 
simulations \cite{geant} of the signal decay modes, \BB\ backgrounds, 
and detector response.
We reconstruct $B$ candidates by combining \KS\ and $\omega$ candidates.
We select $\KS\to\pi^+\pi^-$ decays by requiring the $\pip\pim$ invariant mass 
to be within 12 \mev\ of the nominal \KS\ mass.  We further require the 
three-dimensional flight distance from the beam spot to be greater than three 
times its uncertainty in a fit that requires consistency (fit
probability greater than 0.001) between the flight 
and momentum directions.  We reconstruct $\omega$ mesons through the primary 
\omtoppp\ decay channel from two charged tracks and a \piz\ candidate formed 
from pairs of photons with energy greater than 50 \mev\ and invariant mass 
between 120 and 150 \mev.  The $\pip\pim\piz$
invariant mass is required to be between 735 and 825 \mev.
For the time-dependent analysis, we require $|\dt|<20$ ps and $\sigdt<2.5$ ps.
We find an average of 1.13 $B$ candidates per event.  We choose the candidate 
with the $\pip\pim\piz$ mass nearest to the nominal $\omega$ mass
\cite{PDG2004}.

For a correctly reconstructed $B$-meson candidate, the mass must equal
the nominal $B$ mass and the reconstructed energy must be equal to one-half 
the center of mass energy.  Thus we characterize a candidate kinematically 
by the energy-substituted mass $\mes=\lbrack{(\half
s+\pvec_0\cdot\pvec_B)^2/E_0^2-\pvec_B^2}\rbrack^\half$ and energy
difference $\DE = E_B^*-\half\sqrt{s}$, where the subscripts $0$ and $B$
refer to the initial \UfourS\ and to the $B$ candidate, respectively,
and the asterisk denotes the \UfourS\ rest frame. The resolution on \DE\
(\mes) is about 30 MeV ($3.0\ \mev$). We require $|\DE|\le0.2$ GeV and
$5.25\le\mes\le5.29\ \gev$, and include both of these observables in the
maximum-likelihood (ML) fit (see Sec. \ref{sec:MLfit}).

\section{Backgrounds}
\label{sec:Bkg}
To reject background from continuum $\epem\ra\qqbar$ events ($q=u,d,s,c$), we 
use the angle $\theta_T$ between the thrust axis of the $B$ candidate 
and that of
the rest of the tracks and neutral clusters in the event, calculated in
the center-of-mass frame.  The distribution of $\cos{\theta_T}$ is
sharply peaked near $\pm1$ for combinations drawn from jet-like $q\bar q$
pairs and is nearly uniform for the isotropic $B$ meson decays; we require
$|\cos{\theta_T}|<0.9$. The remaining continuum background dominates the 
samples.

We use MC simulations of \BzBzb\ and \BpBm\ production and decay
to investigate \BB\ backgrounds.  We estimate that this background comprises 
0.2\% of the fit sample.  Since we estimate from simulation studies that
this background would change the signal yield by less than one event, we
do not include a \BB\ component in the fit.

\section{Maximum Likelihood Fit}
\label{sec:MLfit}
We use two unbinned, multivariate maximum-likelihood fits, one to extract 
signal yields and and one to determine the \CP\-violating parameters.  The 
yield fit does not use the 
tagging and \dt\ information in order to reduce systematic errors from
the \dt\ parameterization (though the yields are in excellent agreement
for both fits).  We use six discriminating variables: \mes, \DE, \dt, 
the $\pip\pim\piz$ invariant mass ($m_\omega$), a Fisher discriminant \xf, and 
$\hel\equiv|\cos{\theta_H}|$. The Fisher discriminant combines five variables: 
the polar angles, with respect to the beam axis in the \UfourS\ frame, of the 
$B$ candidate momentum and of the $B$ thrust axis; the tagging category; and 
the zeroth and second angular moments $L_{0,2}$ of the energy flow about the 
$B$ thrust axis.  The moments are defined by 
$ L_k = \sum_i p_i\times\left|\cos\theta_i\right|^k,$ where $p_i$ is the 
momentum of track or neutral cluster $i$, $\theta_i$ is its angle in the 
\UfourS\ frame with respect to the $B$ thrust axis and the sum excludes the $B$
candidate daughters.  The helicity angle $\theta_H$ is the angle, in the 
$\omega$ rest frame, between the normal to the $\omega$ decay plane and the 
$B$ direction.
For each species $j$ (signal or background) and each tagging category 
$c$, we define a total probability density function (PDF) for event $i$ as
\begin{equation}
{\cal P}_{j,c}^i\equiv{\cal P}_j(\mes^i){\cal  P}_j(\DE^i) {\cal P}_j(\xf^i)
{\cal P}_j(m_\omega^i ){\cal P}_j(\hel^i){\cal P}_j(\dt^i,\sigma_{\dt}^i,c)\,,
\end{equation}
where $\sigma_{\dt}^i$ is the error on \dt\ for the event $i$.  
With $n_j$ defined to be the number of events of species $j$ 
and $f_{j,c}$ the fraction of events of species $j$ for each category $c$,
we write the extended likelihood function for all events belonging to category $c$ as
\begin{equation}
{\cal L}_c = \exp{\Big(-\sum_{j} n_j f_{j,c}\Big)}
\prod_i^{N_c} (n_{\rm sig}f_{{\rm sig},c}{\cal P}_{{\rm sig},c}^{i}
                  +n_{\rm bkg} f_{{\rm bkg},c}{\cal P}_{\rm bkg}^{i}),
\end{equation}
where $N_c$ is the total number of input events  in category $c$. 
The total likelihood function for all categories is given as the
product over the tagging categories.  For the yield-only fits, we integrate
over the tagging categories and the product is over the total number of events
in the sample.

We maximize the likelihood function by varying a set of free parameters: 
$S$; $C$; signal and background yields; background shape of \xf, \DE, \mes,
and $\pip\pim\piz$ mass; the fractions of background events in each tagging
category; and six parameters representing the background \dt\ shape.
We determine the PDF parameters for signal from simulation except for \dt,
where we use the \bflav\ data sample discussed in Sec. \ref{sec:dt}.  For the 
continuum background we use (\mes,\,\DE) sideband data to obtain initial 
values, before applying the fit to data in the signal region.  We parameterize 
each of the functions ${\cal P}_{\rm sig}(\mes),\ {\cal P} _{\rm sig} (\DE_k),
\ { \cal P}_j(\xf),\ {\cal P} _{\rm sig}(m_\omega)$ 
and real $\omega$ component of ${\cal P}_{\rm bkg}(m_\omega)$ with either
a Gaussian, the sum of two Gaussians or an asymmetric Gaussian function
as required to describe the distribution.  Slowly varying distributions
(mass, energy or helicity-angle for continuum background and the
combinatorial background component of ${\cal P}_{\rm bkg}(m_\omega)$) are
represented by linear or quadratic dependencies.  The peaking and combinatorial
components of the background $\pip\pim\piz$ mass spectrum each have their
own $\hel$ shapes.  The continuum background in \mes\ is described by
the function $x\sqrt{1-x^2}\exp{\left[-\xi(1-x^2)\right]}$, with
$x\equiv2\mes/\sqrt{s}$ and $\xi$ as a free parameter.  The background \dt\ 
shape is the sum of a core Gaussian convolved with an exponential function 
and two ``tail" Gaussian functions.   To verify the simulated resolutions in 
\DE\ and \mes, we use large control samples of the decays $\Bm\ra\pim\Dz$ with 
$\Dz\ra K^{-}\pip\piz$, which have a topology similar to that of the signal.  
Where the control data samples reveal differences from MC in \mes\ or \DE,
we shift or scale the resolution function used in the likelihood fits.

Before applying the fitting procedure to the data to extract the signal
yields we subject it to several tests.  Internal consistency is checked
with fits to ensembles of ``experiments" generated by MC from
the PDFs.  From these we establish the number of parameters associated
with the PDF shapes that can be left free in addition to the
yields.  Ensemble distributions of the fitted parameters verify that 
the generated values are reproduced with the expected resolution.  The
ensemble distribution of $\ln{\calL}$ itself provides a reference to
check the goodness of fit of the final measurement once it has been
performed. 

We evaluate biases from our neglect of correlations among discriminating 
variables in the PDFs by fitting ensembles of simulated experiments.  Each
simulated experiment has the same number of events as the data for both
background and signal; background events are generated from the background 
PDFs while signal events are taken from the fully simulated MC samples. We find
a positive bias of $7.3\pm0.5$ events.  Since events from a weighted mixture 
of simulated \BB\ background decays are included, the bias we measure 
includes the effect of the neglect of \BB\ background in the fit.

\section{Results}
\label{sec:Physics}
The results of the fits are shown in Table~ \ref{tab:results}. 
The branching fraction is determined from the fit yield, corrected for
the bias discussed above.
The statistical error on the signal yield is equal to the
change in value that corresponds to an increase of $-2\ln{\cal L}$
by one unit from its minimum.  The significance is equal to the
square root of the difference between the value of $-2\ln{\cal L}$ (with
systematic uncertainties included) for zero signal and the value at its
minimum.  Results from simulated experiments suggest a possible underestimate 
of the fit error on $S$ and $C$.  To account for this effect, the statistical 
uncertainty for $S$ and $C$ have been increased by a factor of 1.07.

\begin{table}[htbp]
\caption{Results from yield and \dt\ fits.}
\label{tab:results}
\begin{center}
\begin{tabular}{lcc}
\dbline
Quantity & Yield fit & \dt\ fit \\
\tbline
Events in ML fit        & 9145          & 8070 \\
Fit signal yield        & $96\pm14$     & $92\pm13$  \\
Efficiency ($\epsilon$) & 0.21          &  $-$  \\
$\prodB$                & 0.31          &  $-$  \\
$\epsilon\times\prodB$  & 0.065         &  $-$  \\
Significance            & \somegaKz     &  $-$  \\
\bfemsix                & \romegaKz     &  $-$  \\
$S$                     &      $-$      & $~0.50^{+0.34}_{-0.38}$  \\
$C$                     &      $-$      & $-0.56^{+0.29}_{-0.27}$  \\
\dbline
\end{tabular}
\end{center}
\end{table}

\begin{figure}[!htbp]
\begin{center}
\includegraphics[angle=0,scale=0.80]{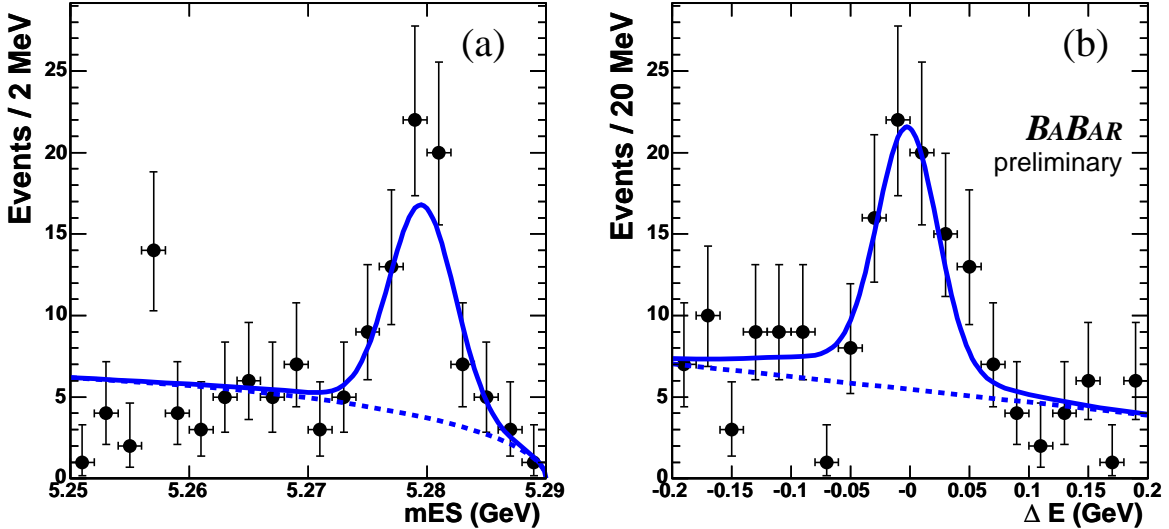}
\caption{ (a) \mes\ and (b) \DE\ projections for \omegaKs\ for data 
subsets optimized from the signal likelihood.
Points with error bars represent the data, solid curves the full fit functions,
and dashed curves the background functions.}
\label{fig:projMbDE}
\end{center}
\end{figure}

\begin{figure}[htbp]
  \begin{center}
  \includegraphics[bb=12 39 475 774 ,scale=0.70]{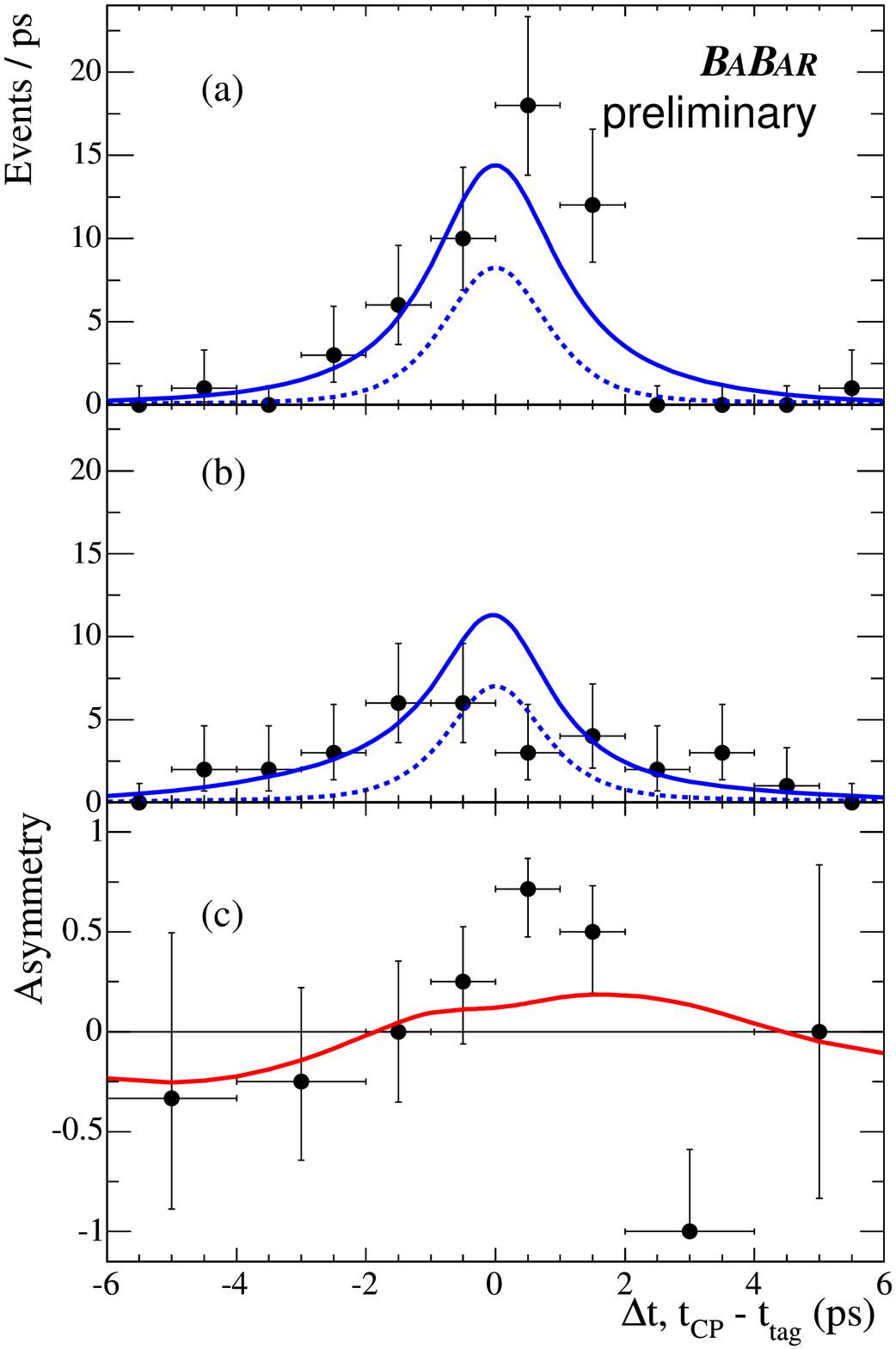}
\end{center}
 \vspace*{.5cm}
 \caption{Projections onto \dt, showing the data (points with errors), fit 
function (solid line), and background function (dashed line), for (a) \Bz\ and 
(b) \Bzb\ tagged events and (c) the asymmetry between \Bz\ and \Bzb\ tags.}
  \label{fig:DeltaTProj}
\end{figure}

In Fig.~\ref{fig:projMbDE}\ we show projections onto \mes\ and \DE\ of a
subset of the data for which the signal likelihood (computed without the 
plotted variable) exceeds a threshold that optimizes the sensitivity.  In 
Fig. \ref{fig:DeltaTProj}, we show plots of the \dt\ distribution for
\Bz- and \Bzb-tagged events and their difference.

\section{Systematic Uncertainties and Crosschecks}
\label{sec:Systematics}

\subsection{Branching Fraction Fit}
Most of the systematic uncertainties arising from lack of knowledge of the
PDFs have been included in the statistical errors since most background
parameters are free in the fits.  For the signal the uncertainties in PDF
parameters are estimated from the consistency of fits to MC and data in
control modes.  Varying the signal PDF parameters within these errors,
we estimate the uncertainties in the signal PDFs to be 0.7 events.
The uncertainty in the fit bias correction is conservatively taken to be 
half of the correction itself.

The above uncertainties are additive in nature.  There are also multiplicative 
systematic errors that are comparable in size, primarily uncertainties in the
efficiency.  
The latter, found from auxiliary studies, include an uncertainty in the 
absolute efficiencies for tracking (1.4\%), \piz\ reconstruction (3.0\%), 
and \KS\ reconstruction (2.1\%).  Our estimate of the systematic error 
in $B$ counting is 1.1\%.  Published data \cite{PDG2004}\ provide the
uncertainties in the $\omega$ product branching fractions (1\%).
The uncertainty in the efficiency of the \costhr\ requirement is 0.5\%.
The total systematic error on the branching fraction is $0.4\timesix$.

\subsection{\dt\ Fit}
The contributions to the systematic uncertainties in $S$ and $C$ are 
summarized in Table~\ref{tab:systtab}.  We evaluate the uncertainties 
associated with the PDF shapes by variation of the parameters describing each 
discriminating variable that is not free in the fit. Systematic errors 
associated with signal parameters (\dt\ resolution function, tagging fractions,
and dilutions) are determined by varying their values within errors.
Uncertainties due to \deltamd\ and $\tau_B$ are obtained 
by varying these parameters by the uncertainty in their
world average values \cite{PDG2004}. All changes are combined in quadrature
obtaining an error of 0.01 for both $S$ and $C$.

We vary the SVT alignment parameters in the signal MC events by the 
size of misalignments found in the real data.  The resulting uncertainty
in both $S$ and $C$ is negligible.

For some tag-side $B$ decays, there is interference between the CKM-suppressed 
$\bar{b}\to\bar{u}c\bar{d}$ amplitude and the favored $b\to c\bar{u}d$
amplitude.  We use simulation, allowing the full range of variation of the 
relevant parameters, to estimate the systematic errors due to this effect 
to be negligible for $S$ and 0.015 for $C$.  An uncertainty 
of 0.02 in $S$ and $C$ is assigned to account for limitations of Monte 
Carlo statistics and modeling of the signal.  The uncertainty in the
effect of the neglect of the small \BB\ background is estimated to be 
less than 0.01 for both $S$ and $C$.  We find that the effects of the
uncertainty in the position and size of the beam spot are negligible.
The total systematic error is obtained by summing individual errors in 
quadrature.

\begin{table}[htbp]
\caption{Estimates of systematic errors.}
\label{tab:systtab}
\begin{center}
\begin{tabular}{lcc}
\hline\hline
Source of error &  $\sigma(S)$ & $\sigma(C)$  \\
\hline
PDF Shapes              &$0.01$  &$0.01$     \\
Tag-side interference   &$0.00$  &$0.02$ \\
\dt\ modeling  		&$0.02$  &$0.02$    \\
\BB\ background		&$0.01$  &$0.01$    \\
\hline
Total                   &$0.02$  &$0.03$    \\
\hline\hline
\end{tabular}
\end{center}
\end{table}

When we fit with the value for $C$ fixed to zero, we find a shift in $S$ of 
0.09, consistent with the correlation of $\sim$20\% between these variables.
We produce samples of pseudo-experiments generated with events produced to 
match the PDF distributions.  From these samples, we verify that the fit bias 
on $S$ and $C$ is negligible and that there is a good agreement between 
expected and observed errors.

\section{Conclusion}
We use a reconstructed signal sample of 96 \omegaKs\ events
to determine the branching fraction, $\BomegaKz=\RomegaKz$.  This value
is in good agreement with previous measurements and with the
world-average value for the charged mode $\BomegaKp=5.1\pm0.7$ \cite{HFAG}. 
This result is also in excellent agreement with the QCD factorization
prediction modified by including final-state interactions \cite{CCS} and
with results from flavor-SU(3)
fits to data for charmless $B$ decays to PV final states \cite{Chiang}.

We also measure the time-dependent CP-violating parameters to be $S=\SomegaKz$
and $C=\ComegaKz$.  The measurement of $C$ is consistent with zero 
and $S$ is in good agreement with the value of $\sin2\beta$ as measured in 
charmonium \Kz\ decays \cite{s2b}. The value for $S$ is also in agreement with,
but more precise 
than, the Belle measurement $S=0.75\pm0.64^{+0.13}_{-0.16}$ \cite{BelleTD}.

\section{Acknowledgments}
\label{sec:Acknowledgments}


We are grateful for the 
extraordinary contributions of our \pep2\ colleagues in
achieving the excellent luminosity and machine conditions
that have made this work possible.
The success of this project also relies critically on the 
expertise and dedication of the computing organizations that 
support \babar.
The collaborating institutions wish to thank 
SLAC for its support and the kind hospitality extended to them. 
This work is supported by the
US Department of Energy
and National Science Foundation, the
Natural Sciences and Engineering Research Council (Canada),
Institute of High Energy Physics (China), the
Commissariat \`a l'Energie Atomique and
Institut National de Physique Nucl\'eaire et de Physique des Particules
(France), the
Bundesministerium f\"ur Bildung und Forschung and
Deutsche Forschungsgemeinschaft
(Germany), the
Istituto Nazionale di Fisica Nucleare (Italy),
the Foundation for Fundamental Research on Matter (The Netherlands),
the Research Council of Norway, the
Ministry of Science and Technology of the Russian Federation, and the
Particle Physics and Astronomy Research Council (United Kingdom). 
Individuals have received support from 
CONACyT (Mexico),
the A. P. Sloan Foundation, 
the Research Corporation,
and the Alexander von Humboldt Foundation.

\end{document}